\documentclass[12pt]{article}
\usepackage{epsf}
\usepackage{graphicx}
\begin{document}
\thispagestyle{empty}

\begin{flushright}
JLAB-THY-04-255 \\
August 10, 2004 \\ 
\end{flushright}

\vspace{1cm}
\begin{center}
{\Large \bf Advances in Generalized  Parton \\ Distribution Study }
\\[1cm]
\end{center}
\begin{center}
{
A. V. RADYUSHKIN\footnote{Also at Laboratory of Theoretical Physics,JINR, Dubna, Russia}
} \\[2mm]  
       {\em Physics Department, Old Dominion University, Norfolk, VA 23529,
USA,} \\[2mm] {\em    Theory Group, Jefferson Lab, Newport News, VA 23606, USA}
\end{center}
\vspace{1cm}

\begin{abstract}
\noindent The basic properties of   generalized parton distributions (GPDs) 
and some recent applications of GPDs are discussed.
\end{abstract}

\vspace{3cm} 

\centerline{\it Talk given at the Workshop ``Continuous Advances in QCD 2004'', 
   Minneapolis,    May 13-16, 2004}
\newpage 

\section{Introduction}
\label{intro}
 
 The concept of Generalized Parton 
 Distributions \cite{Muller:1998fv,Ji:1996ek,Radyushkin:1996nd} 
  is a  modern  tool to provide a more detailed description
  of  hadronic structure.  The need for GPDs is 
  dictated by 
   the present-day 
 situation  
in  hadron physics, namely:     
$i)$ The fundamental particles 
from which the hadrons  are built are known:   quarks and gluons. 
 $ii)$ Quark-gluon interactions are described by   
 QCD whose   Lagrangian is also  known. 
$iii)$ The knowledge of these first principles
is not sufficient at the moment, and we still  need 
hints from experiment to  understand how QCD works, and we must translate  
 information 
obtained on the hadron level 
into the language  of 
quark and gluonic fields. 

One can   consider  projections   of combinations of quark and gluonic fields
onto hadronic states 
$|P \rangle$ : 
$ \langle  \, 0 \, | \, \bar q_\alpha (z_1) \, q_\beta (z_2) \, 
| \, P \, \rangle $, etc., 
      and  interpret them    as hadronic wave functions.
         In principle, solving  the 
  bound-state  equation $H  | P \, \rangle =  E | P \,\rangle $
   one should get   
   complete
  information about  hadronic structure.
  In practice,  the equation
  involving infinite number of Fock components has never been solved.
  Moreover, the  wave functions are not directly accessible 
  experimentally.
   The way out is to  use  phenomenological functions. 
  Well known examples are form factors, 
   usual parton densities, and distribution amplitudes. 
 The new functions, 
   Generalized Parton Distributions \cite{Muller:1998fv,Ji:1996ek,Radyushkin:1996nd} 
   (for   recent reviews,  
   see \cite{Goeke:2001tz,Diehl:2003ny}),   
  are  hybrids of these ``old'' functions which, in their turn,  are the 
  limiting cases of  the ``new''  ones.
  
\section{Form factors,  usual and nonforward parton densities}
\label{sec:1}

    The nucleon electromagnetic  form factors  measurable  
  through elastic $eN$ scattering (Fig. 1, left)  are defined    
  through the  matrix element  
 \begin{equation}{  \langle \, p'  \, 
 |  \, J^\mu (0)  \, |  \, p \,  \rangle 
 =\bar u (p') \left [\gamma^\mu F_1 (t)  - \frac{r^\nu \sigma^{\mu \nu}}
 {2 m_N} F_2 (t) \right ] u(p)  \ ,
 }\end{equation}
 where $r= p-p', t=r^2$.
The  current  is given by the sum of its 
 flavor components  $J^\mu_a (z) = 
   e_a \bar \psi_a (z) \gamma^\mu \psi_a (z)$, hence,  
     $F_{1,2} (t)= \sum_a  \, e_a
  F_{1,2a} (t)$.  
\begin{figure}[h]
 \begin{center}
  \mbox{\hspace{1cm} \epsfysize=4cm
   \epsfbox{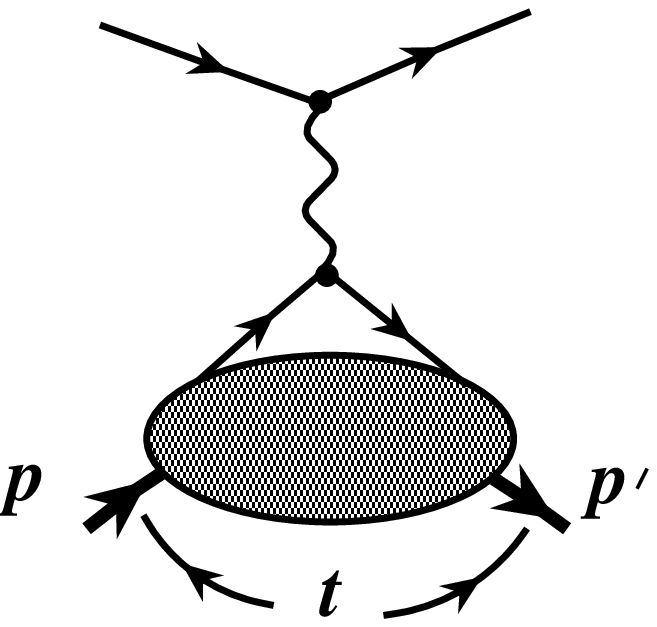} \hspace{1cm}
   \epsfysize=4cm \epsfbox{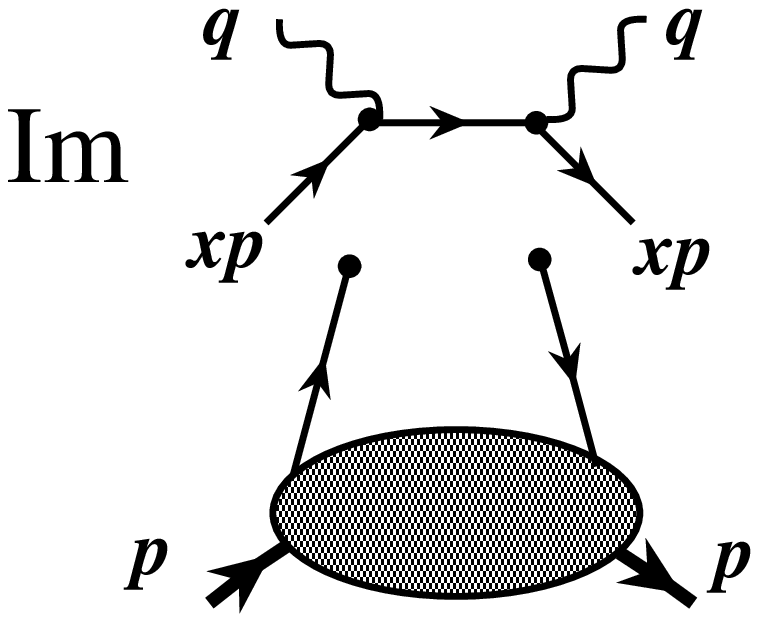}}
   \end{center}
   \caption{Left: Elastic $eN$ scattering  in  one-photon  approximation. 
   Right: Lowest order pQCD factorization for DIS.} 
   \label{fig:1}
    \end{figure}

      The parton densities are defined 
     through forward matrix elements 
  of quark/gluon fields separated by lightlike 
  distances. 
  In  the unpolarized case, 
    \begin{eqnarray}   \langle \, p \, 
 |  \,   \bar \psi_a \left (-\frac{z}{2} \right )  
 \gamma^\mu \psi_a \left (\frac{z}{2}\right ) \, |  \, p \, \rangle
  =2 p^\mu  \int_0^1  
  [ e^{-ix (pz)} f_a (x) - e^{ix (pz)} f_{\bar a} (x)
  ]
 dx  ,
 \label{eq:upd}
 \end{eqnarray}   
  and  
 $f_{a (\bar a)} (x)$ is the probability  to find  $a \, (\bar a)$-quark 
  with momentum $xp$ in a nucleon with momentum $p$. 
One can  access $f_{a (\bar a)} (x)$ through 
deep inelastic scattering (DIS) 
 $\gamma^* N \to X$. 
 Its cross section is given by 
  imaginary part of the forward virtual Compton scattering amplitude. 
   For large  $Q^2 \equiv -q^2$,
   the 
   perturbative QCD (pQCD) factorization works, and 
   the leading order handbag diagram (Fig. 1, right)     
    measures parton densities 
  at the point $x=x_ {Bj} \equiv {Q^2}/{2(pq)}$. 
Note, that    form factor deals with a 
point vertex instead of a light-like separation for   the parton densities, 
and that $p \neq p'$.

 Let us  now ``hybridize''
 form factors with parton densities
 by  writing   form factor components $F_{1a} (t)$   
 as  integrals 
 over the momentum fraction $x$ 
 \begin{equation}  F_{1a} (t) = \int_0^1  \, 
 \left [ {\mathcal F}_{a}(x,t) -  {\mathcal F}_{\bar a} (x,t) \right ] dx   
 \label{local}
\end{equation}
(see Fig. 2, left).   The nonforward 
 parton densities (NPDs) ${\mathcal F}_{a (\bar a)}(x,t) $, 
 coincide in the forward $t=0$ limit with the usual   densities:    
 $ {\mathcal F}_{a(\bar a)} (x,t=0) = 
 f_{a(\bar a)} (x)$.
  A nontrivial   question is the interplay  
  between $x$ and $t$ dependence. The simplest {  factorized} ansatz   
  $ {\mathcal F}_{a} (x,t) = f_{a}(x) F_1 (t)$ 
  satisfies both the forward constraint  
  and  the  {  local} constraint  (\ref{local}).
 \begin{figure}[t]
 \begin{center}
  \mbox{\epsfysize=4cm
   \epsfbox{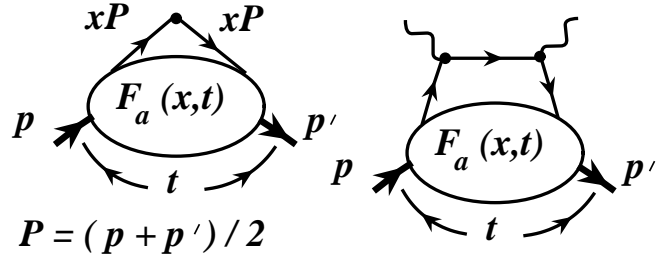}}
   \end{center}
   \vspace{-5mm} 
   \caption{Form factor and  WACS  amplitude in terms of nonforward parton densities.}
   \label{fig:2}
    \end{figure}
    However, using the Gaussian light-cone  wave functions 
{  $\Psi (x_i, k_{i \perp}) \sim 
\exp [-\sum_i k^2_{i \perp}/x_i\lambda^2]$}
suggests \cite{Barone:ej,Radyushkin:1998rt}
   ${\mathcal F}^a(x,t) = f_a(x) e^{\bar x t /2 x \lambda^2 } $.
 Taking $f_a(x)$ from existing parametrizations 
and   $ \lambda^2$   generating  the standard value
 { $\langle k^2_{\perp}\rangle \approx (300 {\rm MeV})^2$}
 for quarks, 
  gives a  reasonable  description \cite{Radyushkin:1998rt} of  $F_1^p(t)$ for  
  $-t \sim 1 -10\,$GeV$^2$.

For small $x$, the usual parton densities have a Regge behavior
$f(x) \sim x^{-\alpha (0)}$. For  $t \neq 0$,  this suggests 
$ {\mathcal F} (x,t) \sim x^{-\alpha (t)}$ or, for a linear Regge trajectory  
  ${\mathcal F}_a (x,t) =  f_a(x) \, x^{-\alpha' t}$.
  With the Regge slope $\alpha' \sim 1$\,GeV$^2$, 
 this model (Fig. 3, dotted lines) allows to obtain correct  charge radii 
 for the proton and neutron \cite{GPRV}.
 At large $t$, the form factor behavior is determined by the
 $x \sim 1$ behavior of $f_a(x)$, giving $t^{-(n+1)}$ if $f_a(x) \sim (1-x)^n$.
 This correlation is  different from the Drell-Yan-West relation,
 which gives $t^{-(n+1)/2}$. One can  conform with  DYW 
 without changing  small-$x$ behavior by  
 taking modified ansatz  ${\mathcal F}_a (x,t) =  f_a(x) x^{-\alpha' (1-x)t}$.
 To apply this   model to  $F_2(t)$,  one needs 
 unknown  {\it magnetic}  parton densities
 $\kappa_a (x)$. 
 To  produce a faster  large-$t$ fall-off 
  of   $F_2(t)$ compared to 
 $F_1(t)$, one can   take  functions  
 $\kappa_a (x)$ having   extra powers of $(1-x)$.
 With  $\kappa_a (x) \sim (1-x)^{\eta_a} f_a(x)$ one   gets
  $F_{2a}(t)/F_{1a}(t)\sim 1/t^{\eta/2}$. 
 The choice \cite{GPRV}   $\eta_u = 1.52, \, \eta_d=0.31$ 
 allows to fit  the JLab polarization 
 transfer data \cite{Gayou:2001qd} on the ratio 
   $F_{2}(t)/F_{1}(t)$
 for the proton, and also provides rather good fits for all
 four nucleon electromagnetic form factors, 
 see solid line curves on Fig. 3.   
  
  \begin{figure}[h]
 \begin{center}
  \mbox{
\epsfysize=7cm
  \epsfbox{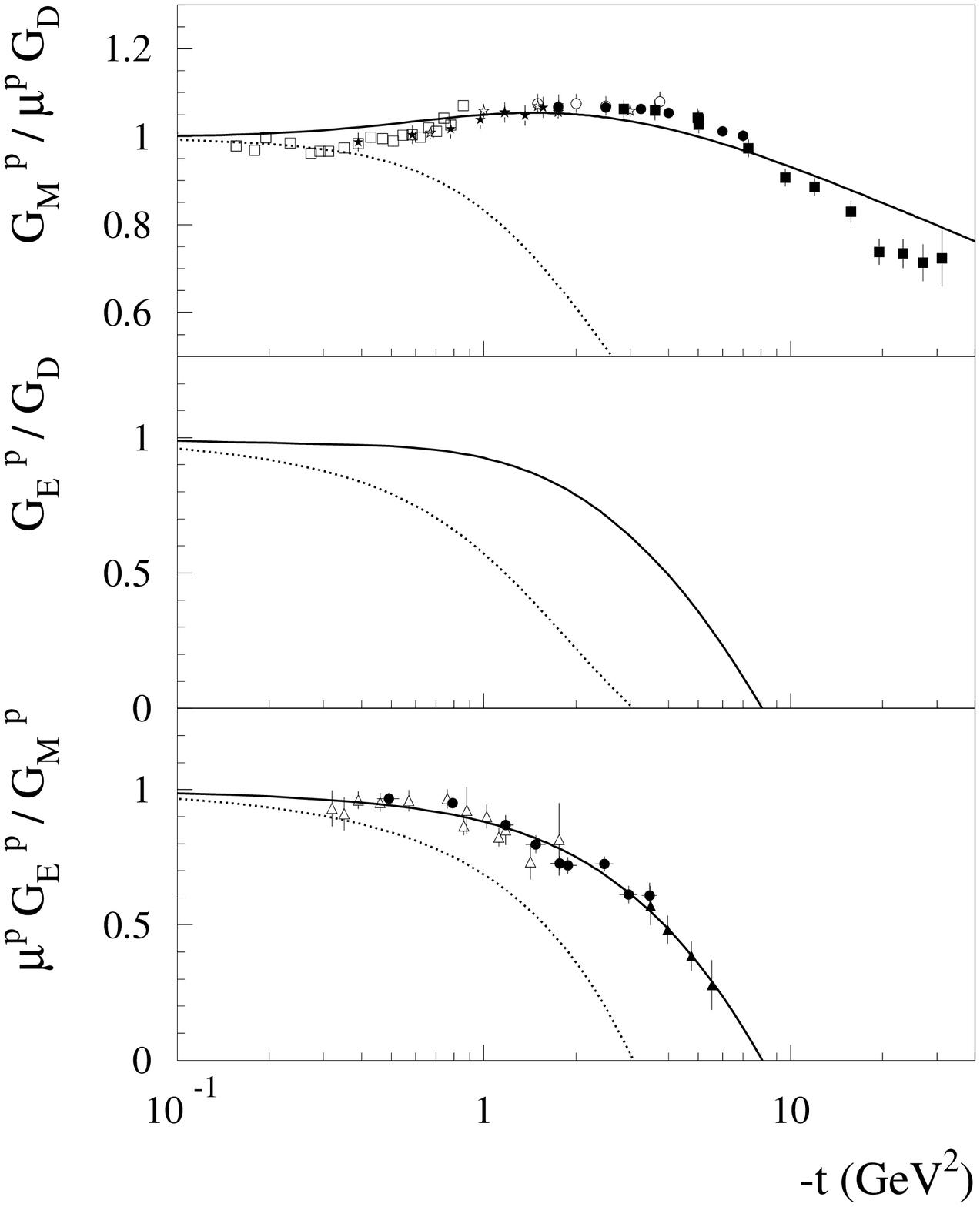} \hspace{1cm} \epsfysize=8cm \epsfbox{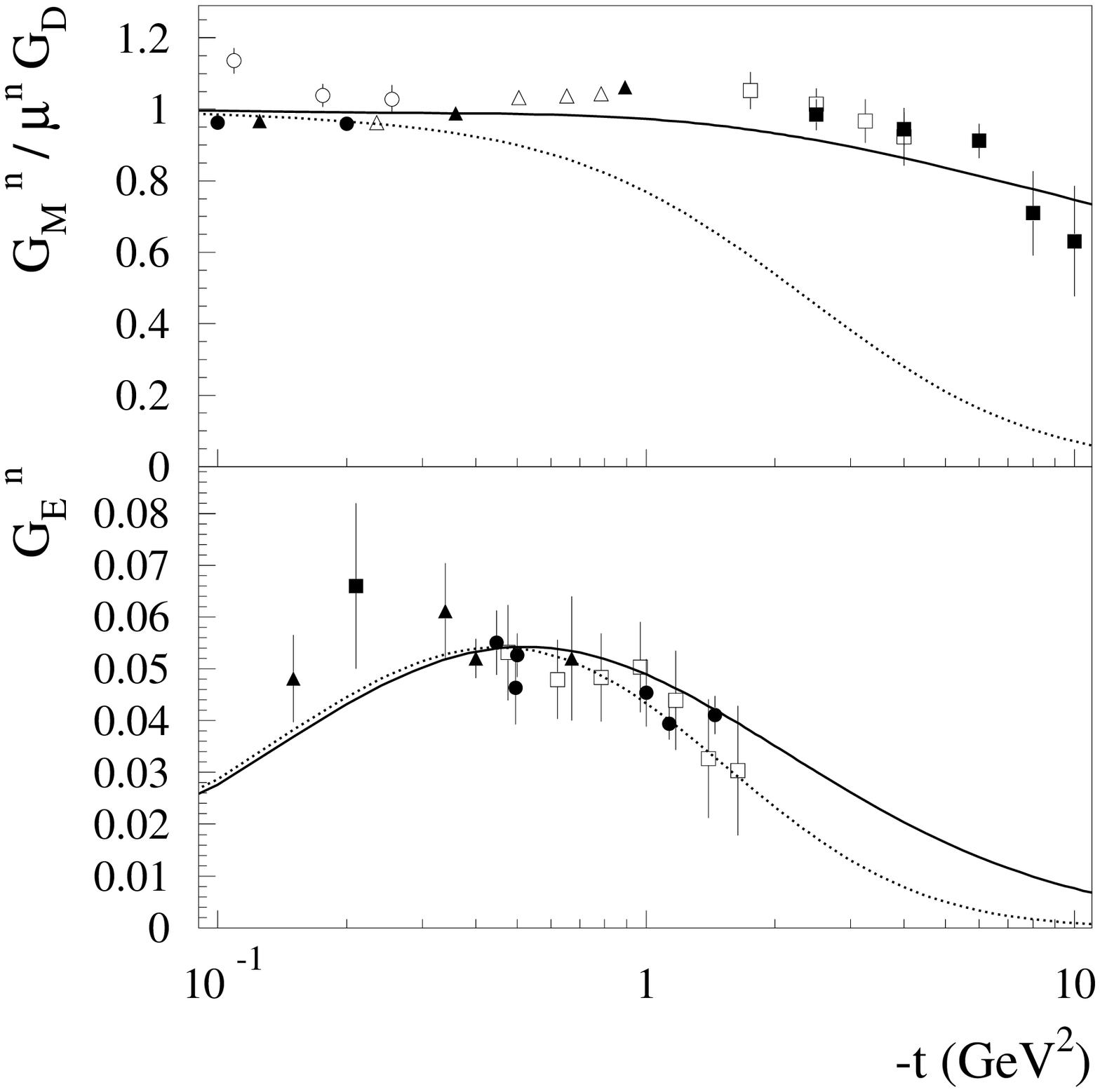}
  }
   \end{center}
   \vspace{-5mm} 
   \caption{Nucleon form factors in Regge-type models for   nonforward parton densities.}
   \label{fig:3}
    \end{figure}

   \section{Wide-angle Compton scattering}

   NPDs also
    appear in   the  wide-angle 
    real Compton scattering (WACS). 
   The handbag term (Fig. 2, right)  is  now 
   given by the $1/x$ moment of ${\mathcal F}^a(x,t)$ and the 
   amplitude  of the Compton scattering  
   off an elementary fermion. The cross  section then can be expressed 
   in terms of  ${\mathcal F}^a(x,t)$ and 
   the  Klein-Nishina (KN) 
 cross section for the Compton scattering  off an electron:
 
    \begin{equation}    \frac{d \sigma}{dt} \sim
    \left [\sum_a e_a^2   \int_0^1  \frac{{\mathcal F}^a(x,t)}{x}
    \, dx  \, \right ]^2 
   \left. \frac{d \sigma}{dt} \right |_{KN}   \  .\end{equation}
 The approach \cite{Radyushkin:1998rt,Diehl:1998kh} based on handbag  dominance  
gives    (with  the Gaussian NPDs fixed from the $F_1(t)$  form factor fitting)   
the results close  both to old Cornell data \cite{Shupe:vg}
and the new preliminary data \cite{ALL,bogdan} of JLab E-99-114
experiment. 
The    predictions  based on pQCD two-gluon hard exchange 
mechanism  depend on the proton wave function and the value of $\alpha_s$.
For the standard choice $\alpha_s =0.3$,  the  pQCD curves 
(see Ref. \cite{Brooks:2000nb}
for the latest calculation)   are  well below the data
even  if one uses extremely asymmetric distribution amplitudes (DAs).
Increasing  $\alpha_s$ to 0.5 gives 
a better agreement, but then  pQCD predictions for
$F_1(t)$ form factor  overshoot the data. 
To remove the   overall  normalization uncertainty,  
 one can consider the ratio $[s^6 d\sigma /dt]/[t^2 F_1(t)]^2$
 sensitive only to the shape of the  proton DA.
 The pQCD results for this ratio presented in Ref. \cite{Brooks:2000nb}
 are an order of magnitude below the data for all DAs considered:
 unlike the  GPD approach, 
 pQCD  cannot simultaneously describe  form factor 
 and WACS cross section data. 
  
  \begin{figure}[h]
 \begin{center}
  \mbox{\hspace{-5mm} \epsfysize=6.5cm
   \epsfbox{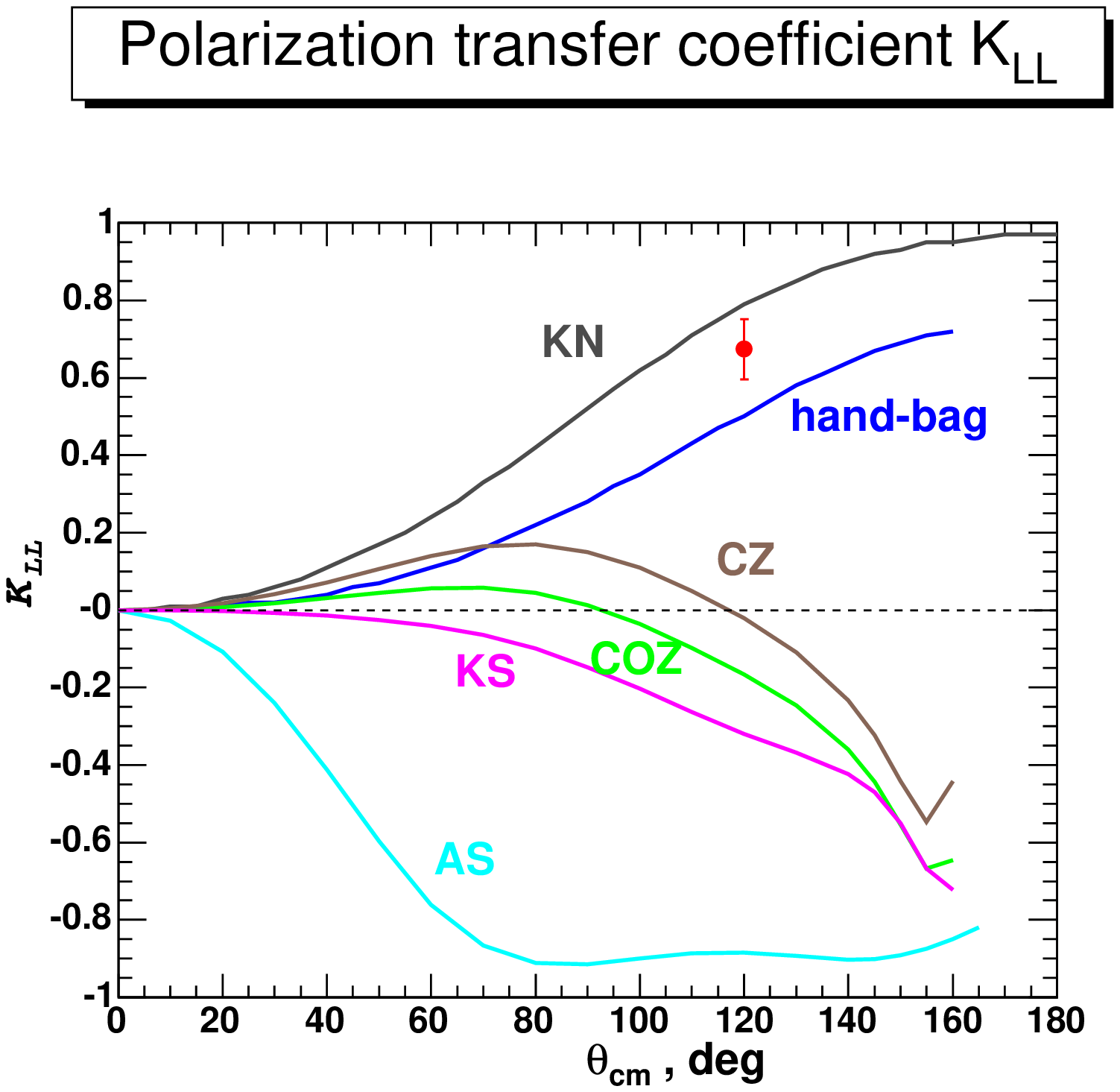} \hspace{-10mm} \epsfysize=6.5cm \epsffile{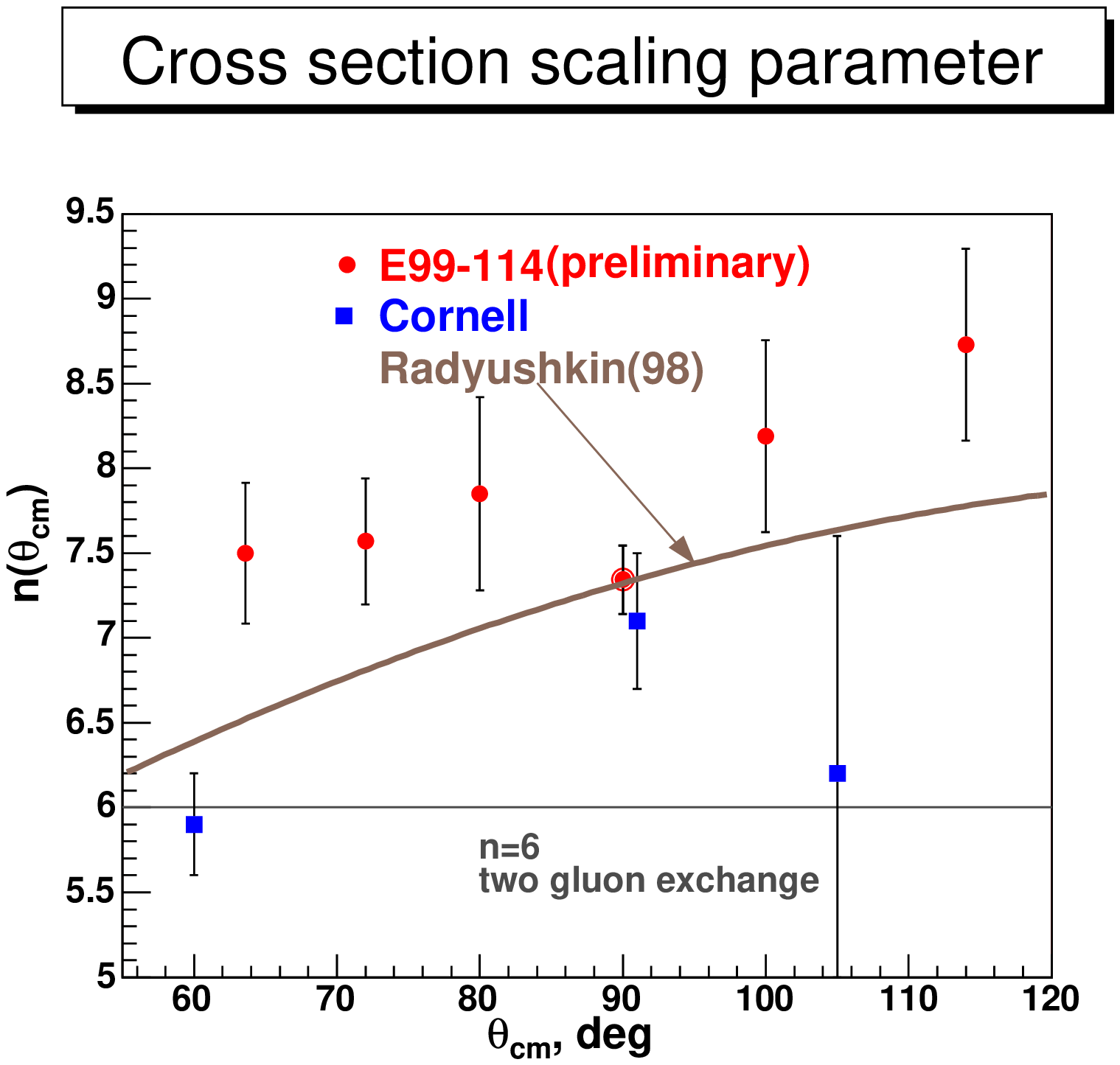}  }
   \end{center}
   \vspace{-5mm} 
   \caption{Comparison of preliminary JLab data with theoretical predictions}
   \label{fig:4}
    \end{figure}

  Furthermore, hard pQCD 
and soft handbag mechanism give drastically different 
predictions \cite{Diehl:1998kh,Brooks:2000nb}  
 for the polarization transfer coefficient $K_{LL}$.
 The  preliminary results (Fig. 4, left) of  E-99-114 experiment \cite{bogdan} 
strongly favor  handbag   mechanism  that  predicts 
a value close to the asymmetry 
for the  Compton scattering on a single free quark. 
  Another ratio-type  prediction of  pQCD  is based on   the
 dimensional  quark counting rules, 
 which give   for WACS $d \sigma / dt \sim s^{-n} f(\theta_{CM})$ 
 with  $n=6$ for all center-of-mass angles $\theta_{CM}$.  The handbag   mechanism
 corresponds to  a power $n$ depending on $\theta_{CM}$,
 in agreement with the preliminary  E-99-114 data \cite{bogdan} (see Fig. 4, right).

   \section{ Distribution amplitudes and pion form factors } 
   
   Distribution amplitudes   describe the hadron structure 
  in situations when pQCD factorization is applicable
  for exclusive processes.
   They are defined through 
    matrix elements   $\langle 0 | \ldots  |p \rangle $
    of light cone  operators.
  For  the pion,  
   \begin{eqnarray} 
    \langle \, 0 \, 
 |  \,   \bar \psi_d (-z/2)\gamma_5  \gamma^\mu 
 \psi_u (z/2) \, |  \, \pi^+(p) \, \rangle
  = i p^\mu f_\pi \int_{-1}^1  e^{-i\alpha (pz)/2}  \varphi_\pi (\alpha) \,  
 d \alpha  
 \,  , \end{eqnarray} 
 with  $x_1 =(1+\alpha)/2, \  x_2 =(1-\alpha)/2$
 being the fractions of the pion momentum carried 
 by the quarks. 
 The simplest case  is 
  $\gamma^* \gamma \to \pi^0$ transition.  
  Its large-$Q^2$ behavior 
  is light-cone dominated: there is no competing Feynman-type soft 
  mechanism.  
   The handbag contribution  for  $\gamma^* \gamma \to \pi^0$ (Fig. 5, left) 
   is proportional to
 the $1/(1-\alpha^2)$ moment of  $\varphi_\pi (\alpha)$ which 
 allows for an experimental discrimination between  the two  popular
 models:  
 asymptotic $\varphi_\pi^{as}(\alpha) = 
 \frac34 (1-\alpha^2) $ and Chernyak-Zhitnitsky DA
  $\varphi_\pi^{CZ} (\alpha) = 
   \frac{15}{4} \alpha^2 (1-\alpha^2)$.    
   Comparison with  data  
favors DA close to $\varphi^{as}_\pi (\alpha)$. 
\begin{figure}[h]
 \begin{center}
  \mbox{\epsfysize=4cm
   \epsfbox{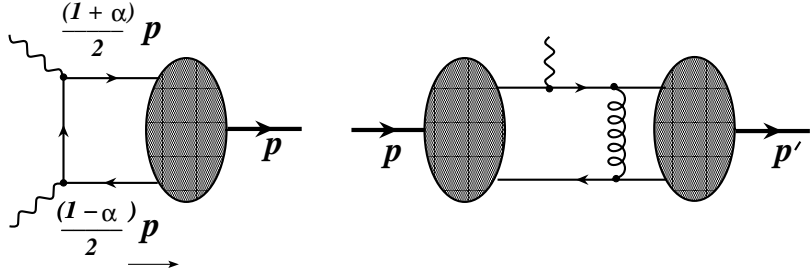}}
   \end{center}
   \vspace{-5mm} 
   \caption{Lowest-order pQCD factorization for $\gamma^* \gamma \to \pi^0$ transition
 amplitude and for the pion EM form factor.}
\label{fig:5} 
    \end{figure}
An important point is  that pQCD works here
 from rather small values  $Q^2 \sim 2\,$GeV$^2$,
 just like in DIS, which is also a purely light-cone dominated process. 
 
 Another  classic application of 
  pQCD to exclusive processes is the  pion electromagnetic 
  form factor.   
    With the asymptotic pion DA, the 
    hard  pQCD
    contribution (Fig. 5, right) to $Q^2 F_{ \pi}(Q^2)$ 
    is  $2\alpha_s /\pi \times 0.7 \,{\rm GeV}^2$,
    less than 1/3 of experimental value which is close 
    to VMD expectation $1/(1/Q^2+1/m_\rho^2)$. 
    The suppression factor $2\alpha_s /\pi$ reflects 
    the usual $\alpha_s /\pi$ per loop penalty 
    for higher-order corrections. The competing soft mechanism  
    is zero order in $\alpha_s$  and  dominates over the pQCD hard term
    at accessible $Q^2$. 
Just like in the case of $F_1^p(t)$, the  soft contribution for $F_{ \pi}(Q^2)$ 
can be modeled 
by nonforward parton densities and easily fits the data 
 (see Ref. \cite{Mukherjee:2002gb}).

\section{   Hard electroproduction processes
and generalized
parton distributions}

A more recent  attempt to use  pQCD
to extract  information 
about hadronic structure is the study of 
deep exclusive photon \cite{Ji:1996ek,Radyushkin:1996nd} or 
meson \cite{Radyushkin:1996nd,Collins:1996fb} electroproduction.
When both $Q^2$ and  $s \equiv (p+q)^2$ are large while  
$t \equiv (p-p')^2$ is small,  
 one can use pQCD factorization of  the amplitudes 
 into a  convolution of a perturbatively calculable short-distance 
 part  and nonperturbative parton functions describing the hadron structure.
The  hard  subprocesses in these two cases have different structure (Fig. 6).
For  deeply virtual Compton scattering (DVCS),  hard  
amplitude has  structure similar to that of the $\gamma^* \gamma \pi^0$
 form factor: the pQCD hard term is of zero order in $\alpha_s$,
 and there is no competing soft contribution.
Thus, we can expect  that pQCD works  from 
 $Q^2 \sim 2\, {\rm GeV}^2$. On the other hand, the deeply virtual 
 meson production process is similar to the pion EM form factor: 
the  hard term has $ O(\alpha_s /\pi) 
 \sim 0.1$ suppression factor. As a result, the 
dominance of the hard pQCD term  
  may be postponed to $Q^2 \sim 5 - 10  \,{\rm GeV}^2$. 
  \begin{figure}
   \begin{center}
  \mbox{\epsfysize=3cm
   \epsfbox{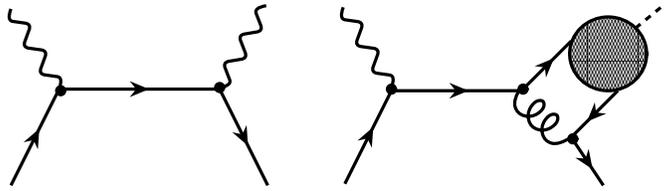}}
   \end{center}
   \vspace{-5mm} 
\caption{Hard subprocesses for 
 deeply virtual photon and meson production.}
\label{fig:6}       
\end{figure}
 Just like in case of pion and nucleon EM form factors,
  the competing soft mechanism 
  can mimic the  power-law $Q^2$-behavior of the hard term.
 Hence, a mere observation of a 
 ``correct'' power   behavior of the cross section 
  is not a proof  that pQCD is already working.
  One should look at several  characteristics of the reaction
  to make 
  conclusions about the reaction mechanism.

 To visualize DVCS's specifics, take  the   $\gamma^* N $ center-of-mass frame, 
 with the initial hadron and the virtual photon moving 
 in opposite directions along the $z$-axis. Since 
 $t$ is small, the hadron and the real photon in the final state
 also move close to the $z$-axis.
This means that the virtual photon momentum $q= q' - x_{Bj}p$  
(where  $x_{Bj}= {Q^2}/{2(pq)}$ is the same  Bjorken variable as in DIS) has 
the component $- x_{Bj}p$  canceled by the momentum transfer $r$. 
In other words,   $r$ has the 
 longitudinal component 
 $r^+= x_{Bj} p^+$, and  
  DVCS has  skewed kinematics:  
 the final hadron's   ``plus'' 
 momentum $(1-\zeta)p^+$  is smaller than that of the initial hadron
(for DVCS,  $\zeta = x_{Bj}$). The plus-momenta $Xp^+$ and $(X-\zeta)p^+$ 
of  the initial and final  quarks
in DVCS are also not equal. 
Furthermore,  the invariant momentum transfer 
$t$ in DVCS is nonzero.  
Thus, the nonforward parton distributions (NFPDs) 
${\mathcal F}_{\zeta}(X;t)$  describing the hadronic structure 
in DVCS depend on $X$, the  fraction of $p^+$  carried by the initial quark,
on  $\zeta$, the skewness parameter characterizing the
difference between   initial and final hadron  momenta,
and on $t$, the invariant 
momentum transfer. 
In the  forward $r=0$ limit, we have a reduction formula 
 ${\mathcal F}^a_{\zeta=0}(X,t=0) = f_a(X)$  
relating NFPDs with the usual parton densities. 
The nontriviality of this relation is that ${\mathcal F}_{\zeta}(X;t)$
appear in the amplitude of the 
exclusive DVCS process, 
 while the usual parton densities are extracted
 from the cross section of the   inclusive  DIS 
reaction.
 In the  limit of  zero skewness, NFPDs  
 correspond to  nonforward parton densities 
  ${\mathcal F}^a_{\zeta=0}(X,t) ={\mathcal F}^a(X,t)$.
 The local limit results in a  formula 
 similar to Eq.(3) : $X$ integral of  
 $  {\mathcal F}^a_{\zeta}(X,t)- {\mathcal F}^{\bar a}_{\zeta}(X,t)$ 
 gives $F_{1a}(t)$. 
 
The NFPD  convention uses  the variables most close
to those of the usual parton densities.
To treat  initial and final hadron momenta  symmetrically, 
Ji  proposed \cite{Ji:1996ek} the  variables in which 
the plus-momenta of the hadrons are 
$(1+\xi)P^+$ and  $(1-\xi)P^+$, and those 
of the active partons are $(x+\xi)P^+$ and $(x-\xi)P^+$, with 
$P= (p+p')/2$ (Fig. 7).
Since $\zeta p^+=r^+=2\xi P^+$, we have   $\xi = \zeta/(2-\zeta)$.
To take into account spin properties of  hadrons 
and  quarks, one needs 4  off-forward parton distributions 
$H, E, \tilde H, \widetilde E$, all  being functions  
 of $x,\xi,t$.
Each OFPD has  3 
distinct regions.
When  $\xi < x < 1$, it is analogous to  usual  quark distributions; when 
$-1 <x< -\xi$,   it is  similar to  antiquark distributions.
In the region 
 $-\xi < x < \xi,$   the ``returning''  quark 
 has  negative plus-momentum, and  should be treated as 
 an outgoing antiquark with momentum $(\xi-x)P^+$. 
 The total $q\bar q$ pair momentum $r^+=2\xi P^+$ is shared by the quarks in 
 fractions $r^+(1+x/\xi)/2 $ and $r^+(1-x/\xi)/2 $.
 Hence OFPD in this region $-\xi < x < \xi $ 
 is similar to a distribution amplitude $\Phi (\alpha)$
 with $\alpha = x/\xi$.
  In the local limit, OFPDs  reduce to form factors
  \begin{figure}[t] \begin{center}
  \mbox{
 \epsfysize=4cm
  \epsfbox{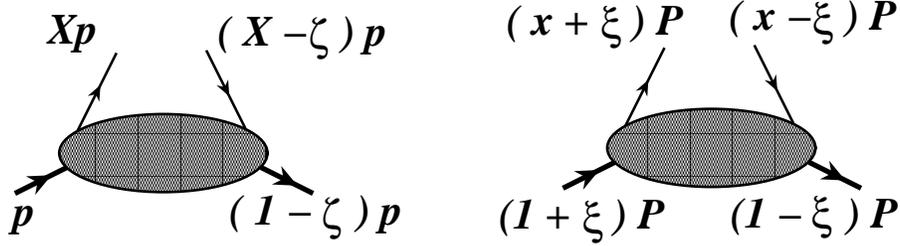} } 
  \end{center} 
 \caption{Comparison of NFPDs and OFPDs. } 
 \label{fig:7}   
 \end{figure} 
\begin{eqnarray}  \sum_a e_a \int \limits_{-1}^1   
 {H}^a (x, \xi;t)
 \,  dx  =F_1(t) \  \ ,  \  \sum_a e_a \int \limits_{-1}^1   
 {E}^a (x, \xi;t)
 \,  dx  =F_2(t) \ . \end{eqnarray}  
The  $E$ function, like $F_2$,   comes with  the $r_{\mu} $ factor, hence,  it is 
 invisible
in DIS  
described 
by  exactly  forward $r=0$ Compton
amplitude. However, the 
 limit   
$E^{a,\bar a} (x, \xi =0 ;t=0)\equiv \kappa^{a,\bar a} (x)$ exists. 
These functions give  the  proton anomalous magnetic moment
$\kappa_p$, and, through    Ji's sum rule \cite{Ji:1996ek},  
the total quark  contribution $J_q$ into 
  the proton spin
\begin{eqnarray} 
\kappa_p =  \sum_a e_a \int \limits_{0}^1   
 (\kappa^a(x) -   \kappa^{\bar a}(x))
 \,  dx  \ ,  \\  J_q = \frac12 \sum_a  \int \limits_{0}^1   
x \,  [f^a(x) +f^{\bar a}(x)+ \kappa^a(x) + \kappa^{\bar a}(x)]  
 \,  dx  \  . \  \  \end{eqnarray}  
 Only valence quarks contribute to
$\kappa_p$, while $J_q$ involves also sea quarks. 
The  determination of the $\kappa$-contribution to Ji's sum rule
is one of the original motivations \cite{Ji:1996ek} to study the GPDs.

\section{ Double distributions} 

 To model GPDs, two approaches are used:  a 
direct calculation 
in  specific  dynamical models 
(bag model, chiral soliton model, 
light-cone formalism, etc.) 
and  phenomenological construction
based on the 
relation of SPDs to usual parton densities 
$f_a(x), \Delta f_a(x)$   and form factors
$F_1(t), F_2(t), G_A(t), G_P(t)$.
The key question 
is the interplay between $x,\xi$ and $t$ 
dependencies of GPDs.  There are not so many cases
in which the pattern of the  interplay is 
evident. 
One example is 
 the function  
$\widetilde E(x,\xi;t)$ that is related to $ G_P(t)$  form factor 
and is dominated 
for small $t$ by the pion pole term $1/(t-m_{\pi}^2)$.  It 
is also proportional to the pion distribution amplitude 
$\varphi (\alpha) \approx \frac34  f_{\pi} (1-\alpha^2)$
taken at $\alpha = x/\xi$. 
  The   construction  of   self-consistent  models for other GPDs
 is performed using   the 
formalism of  double distributions \cite{Muller:1998fv,Radyushkin:1996nd,Radyushkin:1998es}. 

 The main idea behind the double distributions
 is a ``superposition'' 
of $P^+$ and $r^+$ momentum fluxes,
i.e., the representation of the parton 
momentum $k^+ = \beta P^+ + (1+\alpha) r^+/2$  
as the sum of a component $\beta P^+$ due to 
the average hadron momentum $P$ (flowing in the $s$-channel)
and a component $(1+\alpha) r^+/2$ due to the $t$-channel
momentum $r$.  Thus, the double distribution $f(\beta,\alpha)$ 
(we  consider here for simplicity the $t=0$ limit) 
looks like a usual parton density with respect to $\beta$
and like a distribution amplitude with respect to $\alpha$ (Fig. 8).
Using $r^+ = 2 \xi P^+$ gives the connection 
$ x = \beta + \xi \alpha $ between  DD variables $\beta, \alpha$
and   OFPD variables $x,\xi$. 

\begin{figure}[htb]
\centerline{\mbox{  
 \epsfysize=3.5cm
  \epsfbox{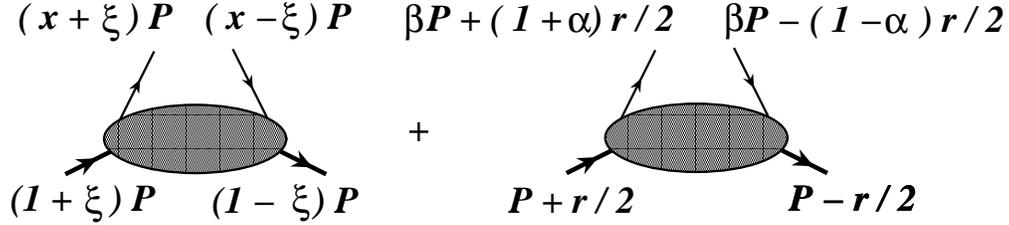}}}
   \caption{Comparison of GPD and DD descriptions.}\label{fig:8} 
\end{figure}

The forward limit $\xi = 0, t=0$ corresponds to $ x = \beta $,
and  gives  the relation between DDs and the 
usual parton densities
 \begin{equation}  \int_{-1 + |\beta |}^{1-|\beta |}  
 f_a(\beta ,\alpha;t=0) \, d \alpha =  f_a(\beta )
\ .
 \end{equation} 
 The DDs live on  the { rhombus} $ |\alpha| + |\beta| \leq 1$
and they are symmetric functions of the ``DA'' variable $\alpha$:  
$f_a(\beta ,\alpha;t) = f_a(\beta ,-\alpha;t)$
(``Munich'' symmetry \cite{Mankiewicz:1997uy}). 
These restrictions suggest a factorized representation  for a DD in the form of a 
product of a
usual  parton density
in the $\beta$-direction and a distribution 
 amplitude 
in the $\alpha$-direction. In particular, a  toy model for a double distribution
$$f(\beta, \alpha) = 3 [(1-|\beta|)^2 - \alpha^2]
   \, \theta( | \alpha |+ | \beta | \leq 1)$$
   corresponds to the toy ``forward''  distribution  
  $f(\beta) = 4(1-| \beta |)^3$, 
 and the $\alpha$-profile like that of the asymptotic pion distribution amplitude.

To get usual parton densities from DDs, one should integrate (scan) them
over vertical lines $\beta = x = {\rm const}$. 
To get OFPDs $H(x,\xi)$ with nonzero $\xi$ from DDs $f(\beta,\alpha)$, 
one should integrate (scan) 
DDs along the parallel lines $\alpha = (x-\beta)/\xi$ 
with a $\xi$-dependent slope. One can call this process the 
DD-tomography.   
 The basic feature of OFPDs  $H(x,\xi)$ resulting from  DDs
is that 
for $\xi =0 $  they reduce to usual parton densities,
and for   $\xi =1 $ they have a shape like a meson  
distribution amplitude.
 
  \begin{figure}[ht]
\centerline{\mbox{  \epsfysize=4cm
  \epsfbox{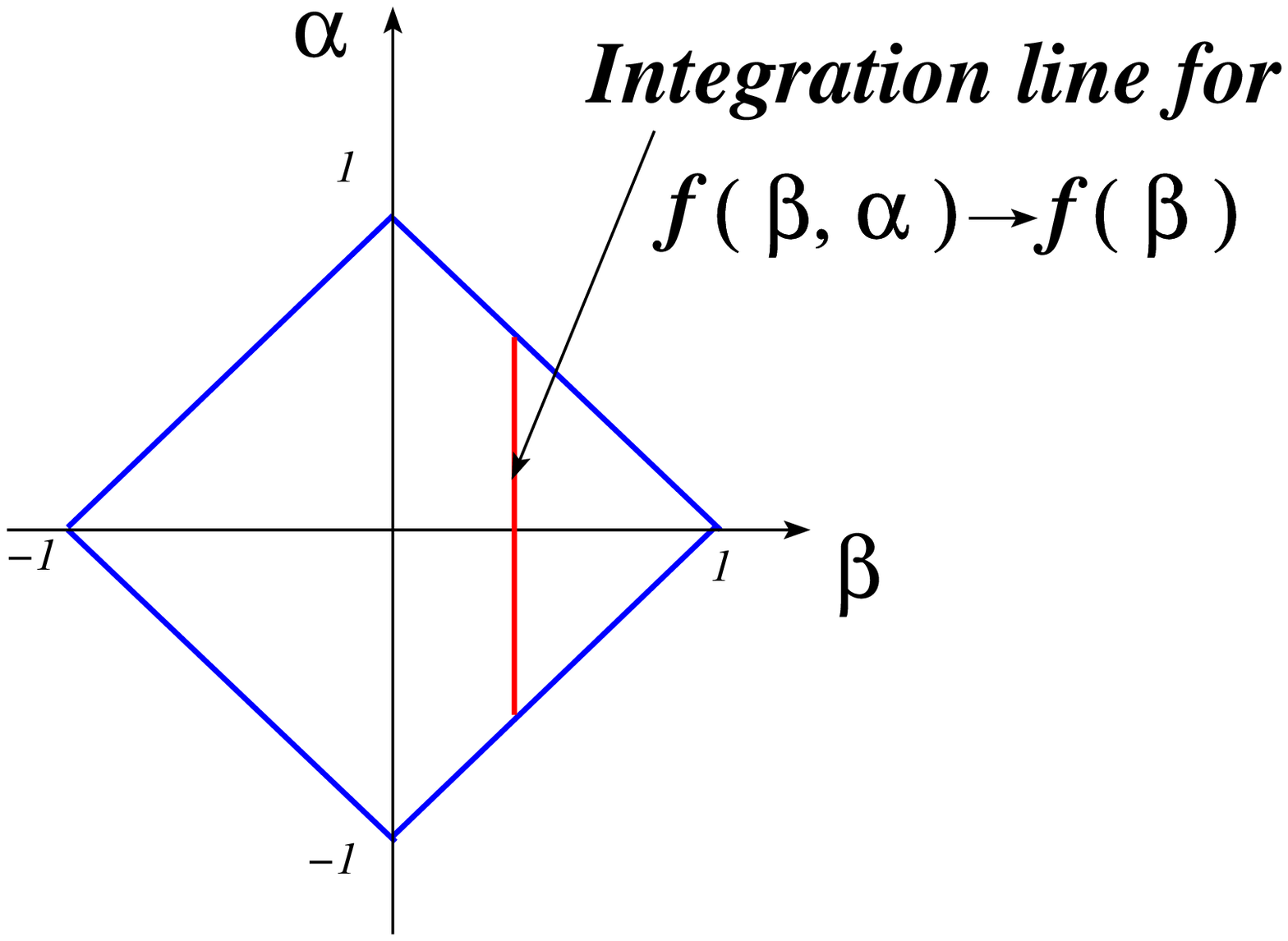}
 \epsfysize=4.5cm
  \epsfbox{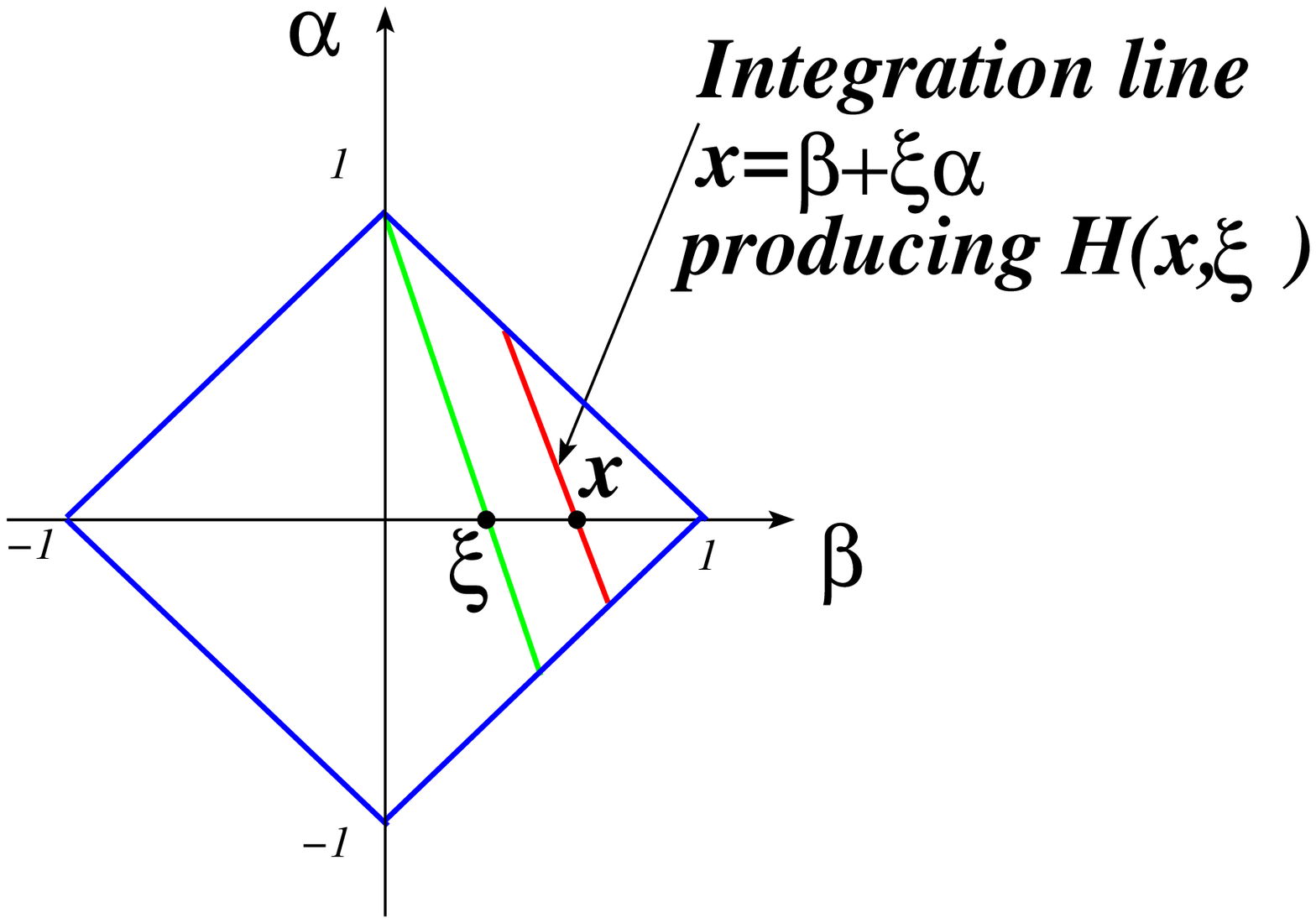}}}
 \caption{Scanning pattern for DD $\to$ SPD conversion}
\label{fig:9}
 \end{figure}

 In fact,  such a  DD modeling 
misses terms proportional to the momentum transfer,
and thus 
invisible in the forward limit. These include  
meson exchange contributions and so-called D-term,
which can be interpreted as $\sigma$-exchange.
The inclusion of the D-term  induces nontrivial
behavior in the central $|x| < \xi$ region
 (for details, see Ref. \cite{Goeke:2001tz}).

 \section{ Conclusions}

Hadronic structure  is a complicated subject, it 
 requires a study from many sides, in many different 
 types of experiments. 
 The description   of specific aspects of 
 hadronic structure 
 is  provided by several different functions:
 form factors, usual parton densities,  
  distribution amplitudes.
Generalized parton distributions  provide a unified description: 
all these functions can  be treated 
as particular or limiting cases  of GPDs $H(x,\xi,t)$. 

{\it Usual Parton Densities}   $f(x)$ correspond to the case
 $\xi = 0, t=0$. 
They describe  a hadron  in terms of  probabilities 
 $\sim |\Psi|^2$. But 
 QCD is a quantum theory: GPDs with $\xi \neq 0$ describe
correlations   $\sim \Psi_1^* \Psi_2$. 
Taking only one point $t=0$ corresponds to integration over impact parameters 
$b_{\perp}$ -  information about the transverse structure is lost. 

{ \it Form Factors} $F(t)$ 
contain  information about the 
 distribution of partons in the transverse plane, but $F(t)$ 
involve  integration over   momentum
fraction $x$ 
 - information about   longitudinal structure is lost.

{\it Nonforward parton densities}.
A simple ``hybridization'' of usual densities
and form factors in terms of NPDs ${\mathcal F}(x,t)$ 
(GPDs with $\xi=0$) 
shows that behavior of $F(t)$ 
is governed   both by transverse 
and longitudinal distributions.
 NPDs provide adequate description of 
nonperturbative soft  mechanism, they  also allow to study
 transition from soft to hard mechanism.
 
 {\it  Distribution Amplitudes}  $\varphi (x)$
provide quantum level information about longitudinal 
structure of hadrons.   
   Information about DAs is  accessible in hard exclusive 
 processes,
 when   asymptotic pQCD  mechanism dominates.  
 GPDs have DA-type structure  in the central region 
$|x| < \xi$.

{\it Generalized Parton Distributions } $H(x,\xi;t)$
 provide a 3-dimensional picture of hadrons.
GPDs also provide some novel  possibilities,
such as   ``magnetic distributions'' related to the
spin-flip GPDs $E_a(x,\xi,t)$.  In particular, the structure of the
nonforward 
density ${\mathcal E}_a (x,t) \equiv E_a(x,\xi=0,t)$
determines the $t$-dependence of $F_2(t)$. 
 Recent JLab data on  the ratio 
  $F_2(t) / F_1 (t)$
 can be   explained   
within  a  GPD-based  model \cite{GPRV} by assuming 
an  extra $(1-x)^{\eta}$ suppression of 
${\mathcal E} (x,t)$.   The forward reductions 
$\kappa^a (x)$  of $E_a(x,\xi,t)$ look as fundamental
as $f^a(x)$  and $\Delta f^a(x)$:
 Ji's sum rule involves $\kappa^a (x)$
on equal footing with $f^a(x)$. 
Magnetic properties of hadrons are strongly sensitive 
to dynamics, thus  providing a   testing ground for models.

A new direction  is  the study of flavor-nondiagonal  distributions:  
 proton-to-neutron GPDs accessible  through   
exclusive charged pion electroproduction process, 
proton-to-$\Lambda$ GPDs (they appear in 
kaon electroproduction);
proton-to-Delta -- this one   
can be related to form factors of  $p\Delta^+$ 
transition (another puzzle for hard pQCD approachable by the NPD model \cite{GPRV}). 
The GPDs for 
$N \to N + {\rm soft} \ \pi$  processes \cite{Goeke:2001tz}   can be used
for testing  the soft
pion theorems and physics of chiral symmetry breaking.

A challenging problem is the separation and flavor decomposition 
of GPDs.  The 
 DVCS amplitude involves all 4 types: $H,E, \widetilde H, \widetilde E$
  of GPDs, so we need  
to study other processes involving different  combinations
of GPDs.  An important observation is that, in hard electroproduction of mesons, the 
spin nature of the produced meson dictates the type
of GPDs involved, e.g.,  for pion electroproduction, only   
$\widetilde H, \widetilde E$ appear, with 
$\widetilde E$  dominated  by the pion pole at small $t$. 
This gives access to 
(generalization of) polarized parton densities 
without polarizing the target.

\section{Acknowledgements} 

I thank the organizers for invitation,  support
and hospitality in Minneapolis.
This work  is supported by the US 
 Department of Energy  contract
DE-AC05-84ER40150 under which the Southeastern
Universities Research Association (SURA)
operates the Thomas Jefferson Accelerator Facility.


\end{document}